\documentclass[usenatbib]{mn2e}

\usepackage{graphicx}
\usepackage{amsmath,amssymb}
\usepackage{hyperref}
\usepackage[varg]{txfonts}
\newcommand{\bbb}{\bigskip}
\newcommand{\be}{\begin{eqnarray}}
\newcommand{\bee}{\begin{enumerate}}
\newcommand{\bit}{\begin{itemize}}
\def\bmat{\begin{pmatrix}}

\def\bkt#1{\left(#1\right)} % normal brackets
\def\bkts#1{\left[#1\right]} % square brackets
\def\bkta#1{\langle#1\rangle} % angle brackets

\def\diff#1#2{{d{#1}\over d{#2}}}
\def\ee{\end{eqnarray}}
\newcommand{\eee}{\end{enumerate}}
\def\eg{\textit{e.g.} }
\newcommand{\eit}{\end{itemize}}
\def\emat{\end{pmatrix}}

\def\ff{\phantom{.}}

\def\iee{\textit{i.e. }}
\newcommand{\ii}{\textit}

\def\lab{\label}

\def\mb#1{\mathbf#1}
\def\mc#1{\mathcal{#1}}

\newcommand{\no}{\noindent}
\newcommand{\nn}{\nonumber}

\def\pr{\prime}

\def\re#1{(\ref{#1})}

\newcommand{\sss}{\smallskip}
\def\sub#1{_{\mbox{\scriptsize{#1}}}}

\def\super#1{^{\mbox{\scriptsize{#1}}}}

\begin{document}
\title[The Effect of Cosmic Magnetic Fields on the Metagalactic Ionization Background]{The Effect of Cosmic Magnetic Fields on the Metagalactic Ionization Background inferred from the Lyman-$\alpha$ Forest  
}

\author[Chongchitnan and Meiksin]{Sirichai Chongchitnan$^{1},$	Avery Meiksin$^{2}$\\
	$^1$ Department of Physics and Mathematics, University of Hull,   Cottingham Rd., Hull, HU6 7RX, United Kingdom.\\
	$^2$ Institute for   Astronomy, University of Edinburgh Royal Observatory, Blackford Hill,   Edinburgh, EH9 3HJ, United Kingdom.
        }

\maketitle

\date{October 2013}

\begin{abstract} 
The sources which reionized the Intergalactic Medium by redshift $\sim6$ are still unknown. A severe constraint on the ionization process is the low emissivity required to maintain the ionization in the Ly$\alpha$  forest. Simulation-calibrated observations suggest a production rate of at most only a few photons per baryon. In this work, we present a new solution to this ``photon-starvation'' problem using a weak background of cosmic magnetic fields, which may be present as a consequence of early-Universe physics and subsequent magneto-hydrodynamical amplification. If present, such magnetic fields can induce density perturbations which are dominant on scales comparable to those probed by measurements of hydrogen-absorption lines at redshifts $z\sim2-5$. We show that a sub-nanoGauss magnetic field, coherent on scale $\sim 1$ Mpc with an almost scale-invariant spectrum, is sufficient to produce significant impact on the effective optical depth, the appearance of the Ly$\alpha$ forest on quasar spectra, the pixel-flux statistics and the  power spectrum of transmitted flux. We also show that such magnetic-field signatures are effectively erased when the metagalactic photoionization rate is increased, hence relaxing the constraint on the cosmic photon budget available for reionization.  
\end{abstract}


\section{Introduction}
\label{sec:Introduction}

Measurements of anisotropies in the cosmic microwave background (CMB) and of  spectra of quasi-stellar objects (QSO) suggest that as early as a few million years after the Big Bang, the primordial hydrogen in the Universe was largely ionized. The process of \ii{cosmic reionization} is thought to have been initiated by high-energy photons produced during the formation of the first luminous objects before a redshift $z\sim11$, with reionization completing as late as $z\sim$ 6 \citep{fanreview, 2012ApJ...756...65Z, 2013ApJS..208...19H, planck}. The details of the reionization process are not yet understood. The galaxies widely believed to be the sources of the reionization have yet to be discovered.

The mystery of the reionization process is compounded by the low ionizing-photon emissivity inferred from the \ii{Ly$\alpha$ forest}, the characteristic fluctuations detected in the spectra of high-redshift quasars arising from the scattering of photons at wavelength $1216(1+z)$ \AA \ff by intervening clouds of neutral hydrogen \citep{meiksinreview}. The ionizing background required to match the measured mean Ly$\alpha$ flux, as calibrated by numerical simulations, corresponds to a source emissivity of only a few ionizing photons per baryon at $z\lesssim6$ \citep{2003ApJ...597...66M, Meiksin05,   2007MNRAS.382..325B}. This suggests the sources that drove reionization had harder spectra than those maintaining the ionization of the Intergalactic Medium (IGM) at $z<6$; or, alternatively, the number of reionizing sources, or the escape fraction of the ionizing photons from the sources, declined sharply as the reionization of the IGM came to completion. Galaxy formation models show the reduction may arise naturally from the suppression of star formation towards the end of the reionization epoch \citep{2012MNRAS.423..862K, 2013MNRAS.429L..94P}, although this scenario still lacks direct observational support.

The constraint on the ionizing photon budget could be relaxed if the density fluctuations in the IGM were sourced by an additional mechanism. In this work, we suggest a solution using a weak background of \ii{cosmic magnetic fields}. It has been widely recognised that a primordial magnetic field would enhance the density fluctuations in the baryonic component \citep{wasserman,   1998PhRvD..58h3502S}, including those giving rise to the Ly$\alpha$ forest \citep{pandey}. The seeds for such fields may have been produced by early-Universe processes such as inflation or symmetry breaking during phase transitions. The seeds are expected to have been amplified by some cosmic dynamo processes (see \cite{kandus, yamazaki, durrer} for recent reviews).

Magnetic fields have been observed to permeate the Universe on a range of physical scales at various magnitudes, from roughly a milliGauss on galactic scales \citep{beck, widrow} down to a microGauss on galaxy-cluster scales \citep{clarke,govoni}. Recent measurements of CMB anisotropies from {\it WMAP}, and most recently from \ii{Planck}, place an upper limit on the magnitude of magnetic fields on cosmological scales ($\sim$ 1 Mpc) of a few nG \citep{paoletti1, planck}. Future CMB-polarization measurements will provide additional constraints on the amplitude of cosmic magnetic fields, since the latter induce Faraday rotation of the plane of polarization of CMB photons \citep{giovannini,kahniashvili3,pogosian}. 

If cosmic magnetic fields are present, they can induce additional density perturbations which are dominant on scales comparable to those probed by measurements of hydrogen-absorption lines at redshifts $z\sim2-5$. There have been only been a few previous studies connecting cosmic magnetic fields to the Ly$\alpha$ forest. \cite{shaw} used SDSS quasar data and a modified CosmoMC code to constrain the magnetic contribution to the matter power spectrum on scale $k\sim 1$ Mpc$^{-1}$. More recently, \cite{pandey} used a semi-analytic approach to generate magnetic-field-induced density fluctuations along lines of sight and constrained the magnetic-field amplitude by comparing with the observations of \cite{faucher2}. 

We extend the previous work to investigate a broader range of signatures of cosmic magnetic fields in the Ly$\alpha$ forest. We find that magnetic-field-induced perturbations increase the fluctuations in the structure of the IGM, deepening and broadening the features while contributing to their longer-wavelength spatial correlations. These features, however, are washed out for high metagalactic photoionization rates. Since the numerical simulations used to calibrate the photoionizing background from the measured mean intergalactic Ly$\alpha$ transmission do not include fluctuations from magnetic fields, they would underestimate the photoionization rate if magnetic-field fluctuations were present. We use an approximate treatment of the structure of the IGM to quantify this effect.

Throughout this work we assume in our fiducial model the following cosmological parameters:\ dimensionless Hubble parameter $h=0.68$ ; present density parameter for matter ($\Omega_m=0.31$), baryon ($\Omega_b=0.048$), radiation ($\Omega_r=2.47\times10^{-5}$), dark energy ($\Omega\sub{DE}=1-\Omega_m-\Omega_r$); scalar spectral index, $n_s=0.96$, normalization of matter power spectrum, $\sigma_8=0.81$, fraction of baryonic mass in helium $Y=0.247$ \citep{planck}.
All magnetic-field amplitudes quoted are comoving:\ $B(0) = B(z)/ (1+z)^2$.

In the next section, we present the formalism used to approximate the Ly$\alpha$ forest, followed by a discussion of the model used for generating the  fluctuations induced by magnetic fields. The effects on the Ly$\alpha$ forest are presented in section \ref{sec:Lya}. We end with a discussion and our conclusions.

\section{The lognormal formalism}
\label{sec:formalism}

Perturbations generated by primordial magnetic fields are dominant on scales on which linear perturbation theory starts to break down, making analytic progress difficult. Here we describe a formalism, due to \cite{bi}, which produces a distribution of mildly non-linear density perturbations at redshift $2\lesssim z\lesssim5$ without the need for \ii{N}-body simulations.
 
Consider the matter overdensity field $\delta(\mb x,z)=(\rho_m(\mb x, z)-\bar\rho_m)/\bar\rho_m$, where $\bar\rho_m$ is the background matter density. Given a distribution of density perturbations in dark matter (generated, say, by inflation), \cite{bi} gave a formalism for calculating the corresponding distribution of \ii{baryonic} matter perturbations based on the hypothesis that the baryonic density perturbations in the mildly nonlinear regime follow a lognormal distribution. In this model, the baryon number density, $n_b$, is given by \be n_b(\mb x , z)&=&n_0(z) \exp\bkt{\delta_b (\mb   x,z) - \bkta{\delta_b^2(\mb x, z)}/2},\lab{logg}\ee
%\be n_b(\mb x , z)&=&n_0(z) {e^{\delta_b (\mb x,z)}\over \bkta{e^{\delta_b (\mb x,z)}}},\lab{logg}\ee
\no where  the background baryon number density $n_0(z)=\bkta{n_b(x,z)}=\Omega_b\rho_c(1+z)^3/(\mu_b m_p)$, and $\mu_bm_p =4m_p/(4-3Y)$ represents mass per baryonic particle. The baryon overdensity, $\delta_b$, is assumed to be a Gaussian random field.  Bi and Davidsen showed that the lognormal approach reproduces the expected distribution of baryons at small and large scales and was shown to be a reasonable approximation when tested against hydrodynamical simulations.

\begin{figure} 
   \centering
   \includegraphics[angle=-90,width=5.5in]{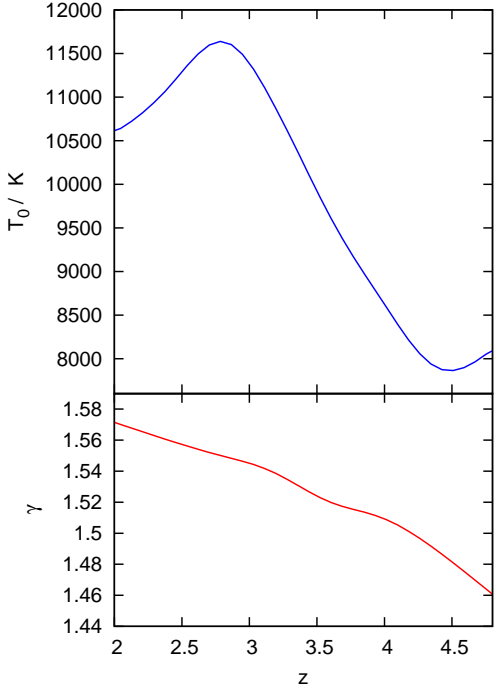} 
   \caption{The IGM temperature at mean density, $T_0(z)$, with varying polytropic index $\gamma$, calibrated using the measurement of \protect\cite{becker}.}
   \label{figgamma}
\end{figure}

Assuming that the IGM temperature evolves smoothly with redshift, the baryon and dark-matter density perturbations on scale $k$ are simply related by the Fourier-space relation
\be \delta_b(\mb k , z) ={\delta_m(\mb k, z)\over 1+x_b^2 k^2}, \lab{filter}\ee
where the comoving Jeans length, $x_b$, is given by \citep{fang}
\be x_b = H_0^{-1}\bkt{2\gamma k T_m(z)\over 3\mu m_p\Omega_m(1+z)}^{1/2}.\ee
Here  $T_m(z)$ is the density-averaged IGM temperature, $\mu=4/(8-5Y)$ is the mean molecular weight of the IGM, and $\gamma$ is the polytropic index defined 
via the equation for the temperature, $T(\mb{x},z)$ of the IGM:
\be T(\mb x, z)= T_0(z) \bkt{n_b(\mb x, z)\over n_0(z)}^{\gamma -1},\ee
where $T_0(z)$ is the temperature at mean density, which we set equal to $T_m(z)$. In contrast with previous works in which  $\gamma$ is usually assumed to be constant, here we use the measurement of  \cite{becker} to infer the values of  $\gamma$ and $T_0$ in the redshift range $2<z<4.8$ as shown in Fig. \ref{figgamma}. Their measurement is derived from 61 high-resolution QSO spectra, and shows $T_0$ increasing from 8000 K at $z=4.4$ to around 12000 K at $z=2.8$. The authors interpreted this rise as coming from the photo-heating of ionized helium (He~II), consistent with He~II reionization at $z\sim3$.

% Figure 1

Inflation and linear perturbation theory predict that the linear matter power spectrum is given by
\be P_m(k,z)&\propto& \mc{M}^2(k,z)k^{n_s-4},\\
\mc{M}(k,z)&\equiv& {2k^2 T(k)D(z)\over 3H_0^2\Omega_m},\lab{mcal}\ee
where the power spectrum is defined via the Fourier autocorrelation
\be \bkta{\delta_m(\mb{k},z), \delta_m(\mb{k^\pr},z)}= (2\pi)^3\delta(\mb k -\mb{k^\pr})P_m(k,z),\ee
$T(k)$ is the transfer function and $D(z)$ the growth factor of density fluctuations. We normalise $P_m(k, z=0)$ so that $\sigma_8$ (square root of the variance of fluctuations on scale $8 h^{-1}$Mpc) equals 0.81. 

From \re{filter}, we see that the 3D baryon power spectrum is given by
\be P_b(k,z)={ P_m(k,z)\over (1+x_b^2k^2)^2}.\lab{3Db}\ee
However, observations of the Ly$\alpha$ forest probe only the one-dimensional baryonic power spectrum along lines of sight. Let 
$$\mc{I}_n(k,z)={1\over 2\pi}\int_k^\infty dk^\pr \bkt{k^\pr}^n P_b(k^\pr,z).$$ The 1D power spectra for the baryon overdensity, $\delta_b(k)$, and velocity perturbations, $v_b(k)$, and their cross-correlation can be expressed as:
\be P_b^{1D}(k,z) &=& \mc{I}_1, \\
 P_v^{1D}(k,z) &=& E^2(z)k^2\mc{I}_{-3},\\
 P_{bv}^{1D}(k,z) &=& i E(z)k \mc{I}_{-1}, \ee
where $E(z)$ determines the growth rate of velocity perturbations and is defined as
$$E(z) \equiv {\diff {\ln D}{\ln a}} \cdot {H(z)\over 1+z}.$$

Bi \& Davidsen's formalism allows the Fourier-space density and velocity fields to be generated given information about their power spectra. To do this, $\delta_b$ and $v_b$  are first decomposed into combinations of uncorrelated Gaussian fields, $u(\mb x)$ and $w(\mb{x})$, using the projection method outlined in \cite{bi1}. The expressions for the independent power spectra $P_u(k,z)$ and $P_w(k,z)$ are
\be P_w(k,z)  &=& \mc I_{-1}^2/\mc I _{-3},\\
P_u(k,z)&=&\mc I_1 -  P_w(k,z). \ee

To obtain the Fourier modes $u(k,z)$ and $w(k,z)$, from their power spectra, we use the polar decomposition
\be u(k,z) = |u|e^{i\phi}, \ee
where $\phi$ is drawn from a uniform distribution $U[0,2\pi]$ and $|u|$ is drawn from the Rayleigh distribution 
\be R(|u|) = {|u|\over \alpha^2} e^{-|u|^2/2\alpha^2}, \quad \alpha^2\equiv P_u(k)/2.\ee
It can be shown that $|u|=\alpha\sqrt{-2\ln \mc{X}}$ where $\mc{X}$ is drawn from another uniform distribution $U[0,1]$. Similarly, $w(k,z)$ can be obtained using independent draws since $u$ and $w$ are uncorrelated Gaussian fields.  
Finally, the density and velocity perturbations of baryons can be written  as
\be \delta_b\super{Inf}(k,z)&=& u(k,z) + w(k,z),\\   
v_b\super{Inf}(k,z) &=& i E(z)kw(k,z) {\mc I_{-3} \over \mc I_{-1}}. \ee
We use the superscript `Inf' to distinguish the inflation-generated perturbations from the magnetic-field induced perturbations, which we now describe.

\section{Perturbations from magnetic fields}
\label{sec:Bfields}

We now consider the baryonic matter perturbations induced by cosmic magnetic fields. The coherent length of such magnetic fields is so large that they may be treated as a stochastic field with homogeneous energy density. The effect of helicity is neglected in this work.

Let $B(\mb x, t)$ be the local amplitude of a homogeneous, non-helical background magnetic field. After recombination, free baryons and magnetic field can be treated as a fluid which can be described by a set of coupled magneto-hydrodynamical equations as shown the pioneering work of \cite{wasserman}. Given a certain amplitude of the magnetic-field component in this fluid, we are interested in the amplitude of the induced density fluctuations in the baryonic component. 

The magnetic-field amplitude scales with the cosmic scale factor, $a(t)$, as $B(\mb x, t)= B(\mb x)/a^2(t)$. This means that the average energy density in magnetic field, $\rho_B\equiv\bkta{B^2(\mb x,t)}/8\pi$, scales like radiation:
\be \rho_B(t)= {\bkta{B^2(\mb x)}\over 8\pi a^4(t)} =   {\rho_{B,0}\over a^4(t)}.\ee

In Fourier space, the magnetic-field power spectrum, $P_B(k)$, is defined by the autocorrelation
\be \bkta{B_i(\mb k)B^*_j(\mb k^\pr)}=(2\pi)^3\delta(\mb k - \mb k^\pr){\mc{P}_{ij}(\hat{k})\over2} P_B(k),\ee
where $\mc{P}_{ij}=\delta_{ij}-\hat{k_i}\hat{k_j}$ is a projection tensor. The power spectrum is commonly parametrized as a power-law with a small-scale cutoff
\be P_B(k) = \begin{cases}
Ak^{n_B},\quad  k\leq k_D
\\
0, \ff\qquad k > k_D
\end{cases}
\ee
\no where  $n_B$ is the magnetic spectral index and  $k_D$ is the cut-off scale, below which the energy in the magnetic fields is dissipated by Alfv\`en-wave damping \citep{jedamzik, subra}. 
This form of the spectrum leads to the expression for the expected magnetic-field amplitude
\be \bkta{B^2}={A k_D^{n_B+3}\over 2\pi^2 (n_B+3)},\ee
valid for $n_B>-3.$ When smoothed using a Gaussian window function, $ \exp(-x^2/\lambda^2)$, we obtain the smoothed amplitude $\bkta{B^2}_\lambda$, which is related to the spectral amplitude by
\be A= {\bkt{2\pi}^{n_B+5} \bkta{B^2}_\lambda\over 2\Gamma\bkt{n_B+3\over2} k_\lambda^{n_B+3}}, \ee
where $k_\lambda = 2\pi/\lambda$. We choose $\lambda = 1$ Mpc in this work,  and for convenience, we denote the \ii{rms} amplitude as
\be B_1 \equiv \sqrt{ \bkta{B^2}\sub{$\lambda=$1Mpc}}.\ee
This choice of $\lambda$ leads to the expression for $k_D$ \citep{kahniashvili} 
\be k_D =\bkts{{140\sqrt{h}}\bkt{2\pi}^{(n_B+3)/ 2}\bkt{1 \mbox{ nG} \over B_1}}^{2/(n_B+5)} \mbox{Mpc}^{-1},\ee
where it is assumed that the Alf\'en wave velocity is proportional to the magnetic-field amplitude. The damping scale is typically small: using $B_1=1$ nG and $n_B=-2.99$, $k_D\approx114$ Mpc$^{-1}$ .

The linear magnetic-field contribution to the matter power spectrum has been derived analytically in \cite{kim} and \cite{gopal}. The expression for the magnetic-field-induced matter power spectrum, in unit of $h^{-3}$Mpc$^{3}$, is:
\be P_{m}\super{MF}(k)&=&{t^4\sub{rec}k^3 \over \bkt{4\pi\rho_b a^3(t\sub{rec})}^2}\int_0^{k_D} dq\int_{-1}^1 d\mu{P_B(q) P_B(\alpha)\over \alpha^2}\mc{K}(k,q,\mu),\nn\\
\mc{K}&=&q^3\bkt{2k^2\mu+kq(1-5\mu^2)+2q^2\mu^3},\lab{nasty}\\
\alpha&=&\sqrt{k^2+q^2-2kq\mu},\nn
\ee %% eq A9 (appendix of kim)
where  $\rho_b$ is the present baryon density and $t\sub{rec}\approx0.371$ Myr is the cosmic time at recombination
%$\rho_m=3H_0^2\Omega_m/(8\pi G).$ 
(see also \cite{paoletti, shaw} for alternative treatments). 

Letting $u=q/k$ brings \re{nasty} to a more manageable form
\be P_m\super{MF}(k) &=& C(k) \int_0^{k_D/k} du\int_{-1}^1 d\mu \mc \ff I(u,\mu),\nn\\
C(k) &=&k^7 \bkts{P_B(k)}^2{t^4\sub{rec} \over \bkt{4\pi\rho_b a^3(t\sub{rec})}^2},\\
\mc I (u,\mu) &=& u^{n_B+3}\bkt{1+u^2-2u\mu}^{n_B/2 -1}\bkt{2\mu+u(1-5\mu^2)+2u^2\mu^3}.\nn
\ee
In this form, we find that $C(k) \sim k^{2n_B+7}$ dominates the behaviour of $P_m\super{MF}(k)$, with small deviation represented by the remaining integrals. The latter must be carefully evaluated across the pole at $(u,\mu)=(1,1)$. Finally, we insert the time dependence using another result of \cite{kim}
\be P_{m}\super{MF}(k,t)&=&\mc{T}^2(t)P_{m}\super{MF}(k),\\
\mc{T}(t)&=&{9\over10}\bkt{t\over t\sub{rec}}^{2/3} +{3\over5}\bkt{t\sub{rec}\over t}-{3\over2}.
\ee

In summary, the magnetic-field-induced matter perturbations are determined mainly by two parameters: the magnetic spectral index, $n_B$, and the \ii{rms} amplitude smoothed at $1$ Mpc, $B_1$.

To calculate the magnetic-field contribution to the baryon perturbations, we replace the $P_m(k,z)$ in Eq. \ref{3Db} by $P_m\super{MF}(k,z).$ and again apply the lognormal formalism. This gives the magnetic-field-induced perturbations, $\delta_b\super{MF}(k,z)$ and $v_b\super{MF}(k,z)$. 

At each point a long the line of sight, we evaluate the Fourier-space density and velocity perturbations $\delta_b\super{Inf}, \delta_b\super{MF}, v_b\super{Inf}, v_b\super{MF}$. An inverse Fourier transform produces the real-space perturbations. The condition $\delta_b(-k)=\overline{\delta_b(k)}$ (and similar for $v_b$) is applied to ensure that the real-space perturbations are indeed real. We take the inflationary and magnetic-field-induced perturbations to be correlated, \iee the amplitudes and phases of both types Fourier modes are drawn from the same distributions. Removing this correlation was shown by \cite{pandey} to have little impact on observables.

Finally, the total real-space density perturbations at each point is
\be\delta_b(x,z)=\delta_b\super{Inf}(x,z)+\delta_b\super{MF}(x,z),\ee
and similarly for the velocity perturbations $v_b(x,z)$. The corresponding baryon number density, $n_b(x, z)$, can then be determined using the lognormal ansatz (Eq. \ref{logg}) and the relation
  \be \bkta{\delta_b^2(x,z)}={1\over2\pi^2}\int_0^\infty {d\ln k} \ff k^3 P_b(k,z).\ee

\section{Effects on the Ly$\alpha$ forest}
\label{sec:Lya}
\subsection{Ly$\alpha$ optical depth}
\label{subsec:tau}
%%%%%%%%%%%%%%%%%%%%%%%%
\begin{figure} 
   \centering
   \includegraphics[angle=-90,width=6.9in]{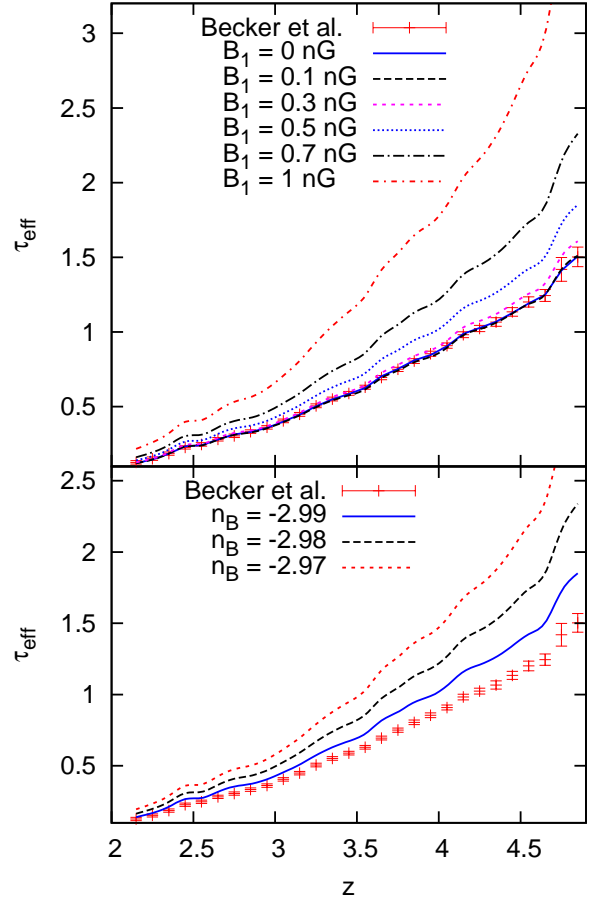} 
   \caption{The effect of cosmic magnetic fields on the effective optical depth, $\tau\sub{eff}$, as a function of redshift. Data points and error bars are from \protect\cite{becker2}. \ii{Upper panel:} the magnitude $B_1$ varies, whilst the magnetic spectral index is fixed at $n_B=-2.99$. \ii{Lower panel:} $n_B$ varies, whilst $B_1=0.5$ nG.  }
   \label{figtau}
\end{figure}
%%%%%%%%%%%%%%%%%%%%%%%%

The optical depth, $\tau(z)$, quantifies the amount of absorption of light emitted at redshift $z$: The intensity of radiation emitted by a QSO is attenuated by a factor of $e^{-\tau(z)}$.   Assuming an approximate Doppler profile for each absorption by the IGM, the optical depth is given by
\be \tau(z) \approx {cI_\alpha \over \sqrt\pi(1+z)}   \int\sub{LOS} \!  {n\sub{HI}(x) \over b(x)} \exp\bkt{ -(\Delta v/b(x))^2}\ff dx \lab{taulos}\ee
where the integration is performed with respect to the comoving distance, $x$, measured towards a point along the line of sight. The Ly$\alpha$ cross section $I_\alpha=4.45\times10^{-18}$cm$^2$, the Doppler parameter $b=\bkt{2k_B T(x,z)/m_p}^{1/2}$ and $\Delta v = v_b + c(z^\pr-z)/(1+z)$ represents the local velocity of the point with redshift $z^\pr$. Averaging the optical depth $\tau(z)$ over the realisations generated by the lognormal approach gives us the \ii{effective} optical depth, 
\be \tau\sub{eff}(z)=-\ln\bkta{e^{-\tau(z)}}, \lab{teff}\ee
which is an observable quantity.

In ionization equilibrium, the number density of neutral hydrogen $n\sub{HI}$ is largely determined by the rate of recombination $\alpha\sub{HI}(T)$ (see \eg \cite{verner}), and the photoionization rate, $\Gamma\sub{HI}(z)\equiv\Gamma_{-12}(z)\times 10^{-12} \mbox{s}^{-1}:$
\be n\sub{HI}={ n_b\ff\alpha\sub{HI}\over \alpha\sub{HI}+\Gamma\sub{HI}(z)/n_e  },\ee
where $n_e$ is the electron number density. We neglect the effect of collisional ionization, which is only significant at temperature $T\gtrsim 10^5$ K \citep{black}.

Fig. \ref{figtau} shows the effective optical depth calculated in the redshift range $2.2\leq z \leq4.8$ using the lognormal approach, with values of $T_0(z)$ and $\gamma(z)$  shown in Fig. \ref{figgamma}. The data points and error bars are from  \cite{becker2} based on 6065 SDSS quasars. We adjust the values of $\Gamma_{-12}$ in each bin so that $\tau\sub{eff}$ matches these data points. The upper panel shows $\tau\sub{eff}$ when the magnetic-field magnitude $B_1$ is increased from $0$ to $1$ nG (with magnetic spectral index $n_B$ fixed at $-2.99$). We observe significant increase beyond the error bars for $B_1\gtrsim 0.3$ nG with most deviation occurring at higher redshifts. This increase stems from the fact that magnetic fields increase  small-scale density fluctuations $\delta_b$, which in turn increase the baryon and the neutral-hydrogen number densities.

The lower panel of Fig. \ref{figtau} shows the variation of $\tau\sub{eff}$ when $n_B$ increases from $-2.99$ to $-2.97$ (with $B_1$ fixed at 0.5 nG). Clearly, there is a strong degeneracy between $n_B$ and $B_1,$ and constraints in this plane have been explored in \cite{yamazaki} and \cite{pandey}. Comparing our results with the latter's, we find similar amplitude of magnetic fields which critically affects $\tau\sub{eff}$, but do not observe the decrease in $\tau\sub{eff}$ for $z>3$ in the presence of magnetic fields as they did. The reason for this is unclear, although it may be partially due to our different models of the IGM\footnote{Pandey \& Sethi  did not vary $T_m$, $\gamma$ and $\Gamma_{-12}$ with redshift, in contrast with our approach.}.

%%%%%%%%%%%%%%%%%%%%%%%%%%%%%%%%
%%%%%%%%%%%%%%%%%%%%%%%%%%%%%%%%
 \begin{figure} 
   \centering
   \includegraphics[angle=-90,width=6.4in]{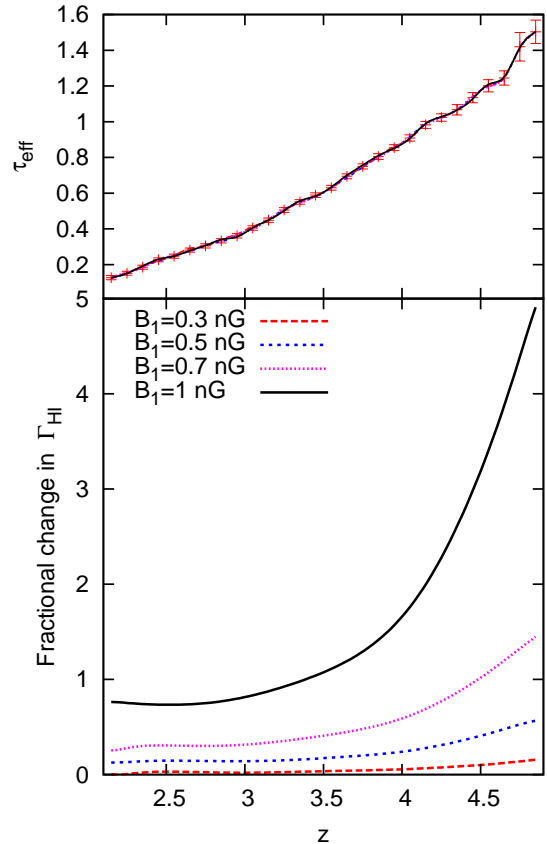} 
   \caption{The degenerate effects of cosmic magnetic fields and the photoionization rate can result in the same effective optical depth, $\tau\sub{eff}$ (top panel -- all four curves overlap). The lower panel shows the fractional increase in $\Gamma\sub{HI}$ required to produce the upper panel, compared to the case with no magnetic fields. }
   \label{ratioG}
\end{figure}
%%%%%%%%%%%%%%%%%%%%%%%%%%%%%%%%
%%%%%%%%%%%%%%%%%%%%%%%%%%%%%%%%

Next, we examine the degeneracy between $B_1$ and $\Gamma\sub{HI}$ by  readjusting $\Gamma\sub{HI}$ in each redshift bin so that $\tau\sub{eff}$ corresponds to the observed values (Fig. \ref{ratioG}, upper panel). The lower panel shows the fractional increase in $\Gamma\sub{HI}$ that would be inferred if magnetic fields with $B_1=0.3-1$ nG are assumed. The fractional increase is defined as 
\be [\Gamma\sub{HI}(B_1\neq0) - \Gamma\sub{HI}(B_1=0)]/ \Gamma\sub{HI}(B_1=0).\ee
We see that the with $B_1=0.5$ nG, for example, photoionization over redshift $3-5$ can be roughly 20--50 percent more efficient when compared to the case without magnetic fields. With $B_1=1$ nG, the increase is more extreme and the photoionization rate can be many times as large towards $z\sim5$.

%%%%%%%%%%%%%%%%%%%%%%%%%%%%%%%%
%%%%%%%%%%%%%%%%%%%%%%%%%%%%%%%%
%%%%%%%%%%%%%%%%%%%%%%%%%%%%%%%%

\begin{figure*} 
   \centering
   \includegraphics[angle=-90,width=7in]{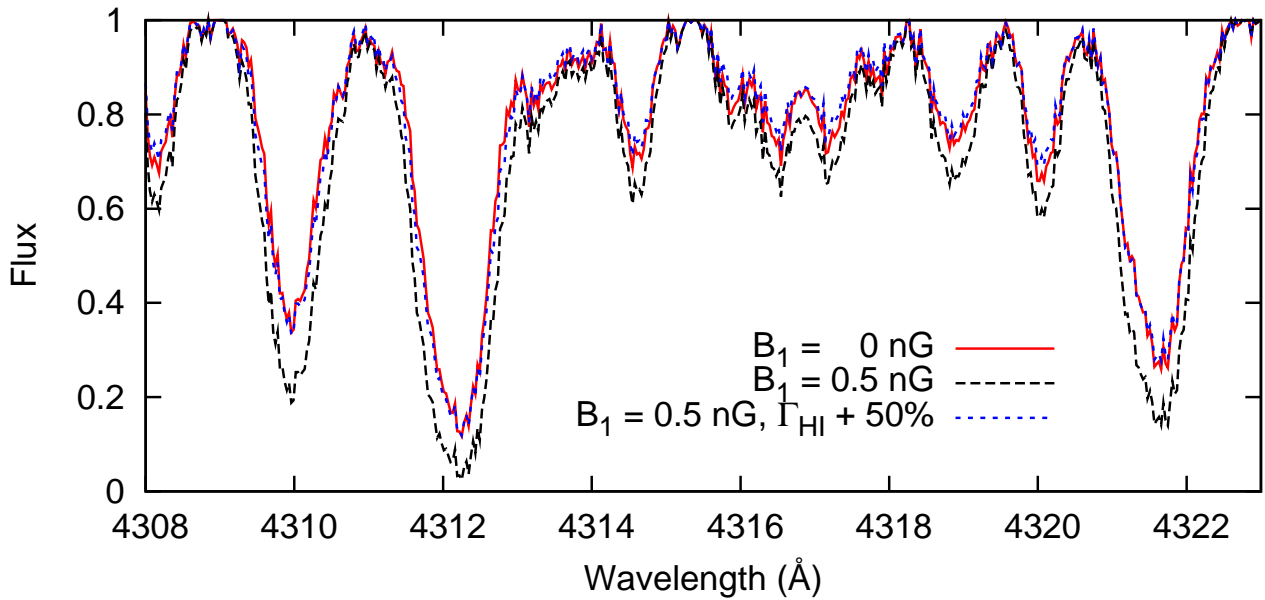} 
   \caption{A synthetic spectrum from $\bar{z}=2.55$ illustrating the effect of cosmic magnetic fields ($B_1=0.5$ nG, $n_B=-2.99$, long-dashed line) on the Ly$\alpha$ forest. The solid line shows the spectrum without magnetic fields. The magnetic fields deepen and broaden the absorption troughs, as discussed in detail the text. These effects, however, can be offset by increasing the background ionization rate. In the Figure,  when  $\Gamma\sub{HI}$ is increased by 50 percent (short-dashed line), the spectrum for $B_1=0.5$ nG is almost identical to the case without magnetic field.}
   \label{synthetic}
\end{figure*}

\subsection{Synthetic spectra}
\label{subsec:spectra}

Next, we use the lognormal approach to produce synthetic QSO spectra and examine the effect of magnetic fields on the Ly$\alpha$ absorption lines in such spectra (see \eg \cite{bi1,bi2,gallerani} for previous work) 

We consider pixels along a line of sight within a box with mean redshift $\bar{z}$. Spectra are typically measured at pixels which are  equally spaced in local-velocity intervals.  The optical depth of a pixel with local velocity $v$ is given by
\be \tau(v) = {cI_\alpha \over \sqrt\pi H(\bar{z})}   \int \!  {n\sub{HI}(v^\pr) \over b(v^\pr)} \exp\bkt{ -[(v-v^\pr-v^\pr_b)/b(v^\pr)]^2}\ff dv^\pr \lab{tauv}\ee
where $v_b$ is the peculiar velocity along the line of sight, and the integration spans all pixels. The flux associated with each pixel is simply $F = e^{-\tau(v)}$. 

An example of such a synthetic flux is shown in Fig. \ref{synthetic} for a QSO at $\bar{z}=2.55$ (similar to Q1017--2046, see \cite{penprase}). The spectrum is drawn from a $10 h^{-1}$Mpc section along the line of sight. To mimic the instrumental profile, the spectrum is further convolved with a Gaussian function with full width at half maximum of 6.7 km/s (roughly the resolution of  HIRES spectrograph) and resampled at velocity interval of 2.1 km/s. Finally, we also add to the flux a Gaussian noise with zero mean and $\sigma\sub{noise}=0.02$. The resulting spectrum is shown in solid line in Fig. \ref{synthetic}.

Cosmic magnetic fields of strength $B_1=0.5$ nG ($n_B=-2.99$) is then added and the spectrum is recalculated. The result is shown in dashed line in Fig. \ref{synthetic}. We clearly see the deepened and broadened absorption troughs which result from additional inhomogeneities from the magnetic fields, consistent with the findings in the previous section.

The degeneracy between $B_1$ and $\Gamma\sub{HI}$ is illustrated by the spectrum in short-dashed line, where $\Gamma\sub{HI}$ is increased by 50 per cent. The resulting spectrum is almost identical to the spectrum without magnetic fields, showing that increasing the photoionization rate can effectively erase the magnetic-field imprints on the Ly$\alpha$ spectrum. This enhancement is slightly greater than that expected from Fig. \ref{ratioG}, since the latter is calculated by averaging over multiple lines of sight without instrumental and noise considerations.

\begin{figure} 
   \centering
   \includegraphics[angle=-90, width=3.4in]{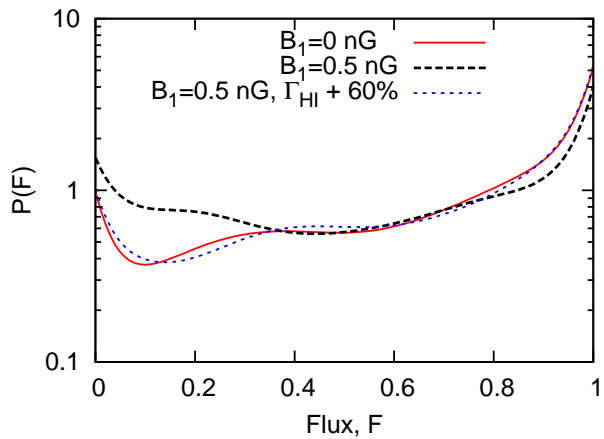} 
   \caption{The normalised probability density distribution of the flux,$P(F)$, from synthetic spectra with $\bar{z}=3$. The presence of magnetic fields with strength $B_1=0.5$ nG (long dashed) skews the pdf towards $F\approx0$, indicating deeper absorption troughs compared to the case without magnetic field (solid line). The magnetic signature can again be masked by enhancing the photoionization rate by 60 percent (short dashed).}
   \label{pdf}
\end{figure}

We further investigate if these degenerate effects on $\tau\sub{eff}$ can be distinguished in the pixel flux statistics $P(F)$. We assume pixel bins of width $\Delta F$=0.05, and normalise the flux probability density distribution, $P(F)$, so that $\sum P(F)\Delta F=1$. Figure \ref{pdf}(solid line) shows such a flux pdf taken from synthetic a spectrum with $\bar{z}=3$, exhibiting the usual double-peak feature (see \eg \cite{becker4,kim2}). A magnetic field with $B_1=0.5$ nG $(n_B=-2.99)$ is again added. This skews the PDF towards the region where $F\approx0$, indicating deeper absorption troughs as expected. As before, we were able to mask the magnetic-field imprints by enhancing the photoionization rate by 60 percent in this case  (short-dashed line), producing an almost identical pdf to the case without magnetic field. 

\begin{figure} 
   \centering
   \includegraphics[ width=3.4in]{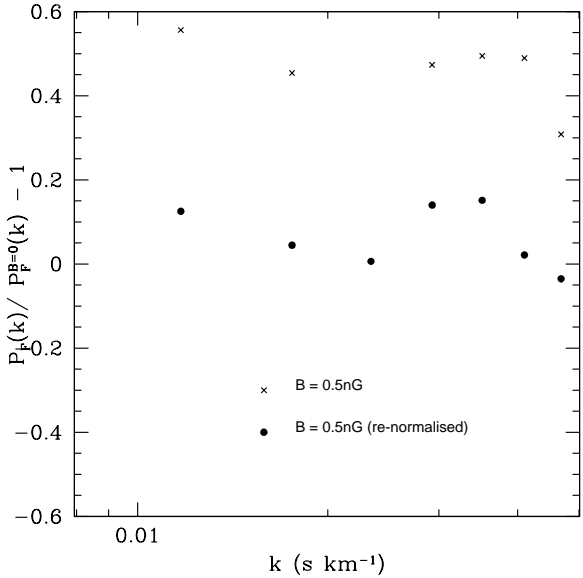} 
   \caption{The relative change in the flux power spectrum allowing for magnetic field fluctuations (derived from the spectrum in Figure \ref{synthetic}). The presence of magnetic fields with strength $B_1=0.5$ nG (crosses) substantially boosts the power. The magnetic signature, however, is largely masked by enhancing the photoionization rate by 50 percent (dots).}
   \label{epfk}
\end{figure}

Adding a magnetic field with $B_1=0.5$ nG $(n_B=-2.99)$ also substantially boosts the flux power spectrum, as shown in Figure \ref{epfk} (derived from the spectrum in Figure \ref{synthetic}). This is consistent with enhanced structure on small scales, as the flux power spectrum integrates along the lines of sight. It will in general also include the effects of larger-wavelength modes. The boost is largely compensated for, however, by increasing the photoionization rate by 50 percent, necessary to recover the mean observed transmission. As \citet{shaw} appear not to have included the photoionizing background as a free parameter, it is unclear how meaningful their Ly$\alpha$ forest constraint on a primordial magnetic field is.

\sss

In summary, this section illustrates that by introducing a sub-nanoGauss amplitude of cosmic magnetic fields, the metagalactic photoionization rate inferred from Ly$\alpha$-forest measurements can be significantly enhanced.

\section{Conclusions and discussion}

We have shown how the ``ionizing-photon budget" problem can be alleviated by introducing a weak background of cosmic magnetic fields. Our main results are summarised below.

Firstly, we showed quantitatively how cosmic magnetic fields induce baryonic density perturbations on top of the standard $\Lambda$CDM perturbations,  giving rise to an enhanced population of Ly$\alpha$ clouds. A weak magnetic field of order $\sim0.1-1$ nG, coherent on scale $\sim 1$ Mpc, with an almost scale-invariant spectrum was shown to be sufficient to produce significant impact on observables derived from high-redshift QSO spectra, including i) the effective optical depth, $\tau\sub{eff}$, of photons to Ly$\alpha$ absorption, ii) the spectra themselves, iii) the pixel-flux statistics, and iv) the flux power spectrum. In all four observables, we found a consistent picture of magnetic fields creating deeper, broader absorption troughs along the lines of sight. The flux power spectrum in general will also include the effects of longer wavelength modes.

Furthermore, we examined the sensitivity of the magnetic-field effects on the above observables to the assumed metagalactic photoionization rate, $\Gamma\sub{HI}(z)$. Decreasing the photoionization rate is degenerate with increasing the fluctuations in the density of neutral hydrogen induced by cosmic magnetic fields. The precise nature of this degeneracy could be explored using a likelihood analysis, but we leave this for future work.

The results in Figs. \ref{ratioG}$-$\ref{epfk} demonstrate that a sub-nanoGauss level of magnetic field from an almost scale-invariant spectrum is sufficient to significantly enhance the required value of $\Gamma\sub{HI}$(z), with the amount increasing with redshift to a factor of several. This would substantially ease the tension between the number of ionizing photons per baryon required to reionize the IGM and the number required to maintain the IGM at its level of ionization afterwards, as inferred from the Ly$\alpha$ forest.

The thermal and radiative properties of the IGM are crucial in our calculations. In particular, we used values of the temperature at mean density, $T_0(z)$, and the polytropic index, $\gamma(z)$, inferred from the measurements of \cite{becker}. More realistically, the presence of large-scale  magnetic fields would alter these parameters due to magnetic energy dissipation. The formation of the first ionized sources are also likely to be affected, meaning that $\Gamma\sub{HI}$ will carry some dependence on the magnetic-field strength. Some of these issues have been investigated in \cite{schleicher2} and \cite{sur}, although much larger hydrodynamical simulations are needed to elucidate the precise relationship between cosmic magnetic fields and the IGM.

Finally, although neglected in this work, cosmic helicity may have played an important role in the amplification of seed fields through the so-called inverse cascade mechanism  \citep{brandenburg} and could also leave novel imprints in the CMB \citep{kahniashvili2}. It will be interesting to investigate the imprints of helical magnetic fields on the IGM in future work. 

\bbb

\no\ii{Acknowledgment}: We are grateful for helpful discussions with Kanhaiya Pandey, and for support from the Carnegie Trust of Scotland.

%\appendix
%\section{MF integral}

\bibliographystyle{mn2e}
\bibliography{pmf2}

\begin{thebibliography}{48}
\expandafter\ifx\csname natexlab\endcsname\relax\def\natexlab#1{#1}\fi

\bibitem[{{Beck} {et~al}\mbox{.}(1996){Beck}, {Brandenburg}, {Moss},
  {Shukurov}, \& {Sokoloff}}]{beck}
{Beck} R., {Brandenburg} A., {Moss} D., {Shukurov} A., {Sokoloff} D., 1996,
  \araa, 34, 155

\bibitem[{{Becker} {et~al}\mbox{.}(2011){Becker}, {Bolton}, {Haehnelt}, \&
  {Sargent}}]{becker}
{Becker} G.~D., {Bolton} J.~S., {Haehnelt} M.~G., {Sargent} W.~L.~W., 2011,
  \mnras, 410, 1096

\bibitem[{{Becker} {et~al}\mbox{.}(2013){Becker}, {Hewett}, {Worseck}, \&
  {Prochaska}}]{becker2}
{Becker} G.~D., {Hewett} P.~C., {Worseck} G., {Prochaska} J.~X., 2013, \mnras,
  430, 2067

\bibitem[{{Becker}, {Rauch} \& {Sargent}(2007){Becker}, {Rauch}, \&
  {Sargent}}]{becker4}
{Becker} G.~D., {Rauch} M., {Sargent} W.~L.~W., 2007, \apj, 662, 72

\bibitem[{{Bi}(1993)}]{bi1}
{Bi} H., 1993, \apj, 405, 479

\bibitem[{{Bi} \& {Davidsen}(1997)}]{bi}
{Bi} H., {Davidsen} A.~F., 1997, \apj, 479, 523

\bibitem[{{Bi}, {Boerner} \& {Chu}(1992){Bi}, {Boerner}, \& {Chu}}]{bi2}
{Bi} H.~G., {Boerner} G., {Chu} Y., 1992, \aap, 266, 1

\bibitem[{{Black}(1981)}]{black}
{Black} J.~H., 1981, \mnras, 197, 553

\bibitem[{{Bolton} \& {Haehnelt}(2007)}]{2007MNRAS.382..325B}
{Bolton} J.~S., {Haehnelt} M.~G., 2007, \mnras, 382, 325

\bibitem[{{Brandenburg}, {Enqvist} \& {Olesen}(1996){Brandenburg}, {Enqvist},
  \& {Olesen}}]{brandenburg}
{Brandenburg} A., {Enqvist} K., {Olesen} P., 1996, \prd, 54, 1291

\bibitem[{{Clarke}, {Kronberg} \& {B{\"o}hringer}(2001){Clarke}, {Kronberg}, \&
  {B{\"o}hringer}}]{clarke}
{Clarke} T.~E., {Kronberg} P.~P., {B{\"o}hringer} H., 2001, \apjl, 547, L111

\bibitem[{{Durrer} \& {Neronov}(2013)}]{durrer}
{Durrer} R., {Neronov} A., 2013, ArXiv e-prints 1303.7121

\bibitem[{{Fan}, {Carilli} \& {Keating}(2006){Fan}, {Carilli}, \&
  {Keating}}]{fanreview}
{Fan} X., {Carilli} C.~L., {Keating} B., 2006, \araa, 44, 415

\bibitem[{{Fang} {et~al}\mbox{.}(1993){Fang}, {Bi}, {Xiang}, \&
  {Boerner}}]{fang}
{Fang} L.-Z., {Bi} H., {Xiang} S., {Boerner} G., 1993, \apj, 413, 477

\bibitem[{{Faucher-Gigu{\`e}re} {et~al}\mbox{.}(2008){Faucher-Gigu{\`e}re},
  {Prochaska}, {Lidz}, {Hernquist}, \& {Zaldarriaga}}]{faucher2}
{Faucher-Gigu{\`e}re} C.-A., {Prochaska} J.~X., {Lidz} A., {Hernquist} L.,
  {Zaldarriaga} M., 2008, \apj, 681, 831

\bibitem[{{Gallerani}, {Choudhury} \& {Ferrara}(2006){Gallerani}, {Choudhury},
  \& {Ferrara}}]{gallerani}
{Gallerani} S., {Choudhury} T.~R., {Ferrara} A., 2006, \mnras, 370, 1401

\bibitem[{{Giovannini} \& {Kunze}(2008)}]{giovannini}
{Giovannini} M., {Kunze} K.~E., 2008, \prd, 78, 023010

\bibitem[{{Gopal} \& {Sethi}(2005)}]{gopal}
{Gopal} R., {Sethi} S.~K., 2005, \prd, 72, 103003

\bibitem[{{Govoni} \& {Feretti}(2004)}]{govoni}
{Govoni} F., {Feretti} L., 2004, International Journal of Modern Physics D, 13,
  1549

\bibitem[{{Hinshaw} {et~al}\mbox{.}(2013){Hinshaw}, {Larson}, {Komatsu},
  {Spergel}, {Bennett}, {Dunkley}, {Nolta}, {Halpern}, {Hill}, {Odegard},
  {Page}, {Smith}, {Weiland}, {Gold}, {Jarosik}, {Kogut}, {Limon}, {Meyer},
  {Tucker}, {Wollack}, \& {Wright}}]{2013ApJS..208...19H}
{Hinshaw} G. {et~al.}, 2013, \apjs, 208, 19

\bibitem[{{Jedamzik}, {Katalini{\'c}} \& {Olinto}(1998){Jedamzik},
  {Katalini{\'c}}, \& {Olinto}}]{jedamzik}
{Jedamzik} K., {Katalini{\'c}} V., {Olinto} A.~V., 1998, \prd, 57, 3264

\bibitem[{{Kahniashvili}, {Maravin} \& {Kosowsky}(2009){Kahniashvili},
  {Maravin}, \& {Kosowsky}}]{kahniashvili3}
{Kahniashvili} T., {Maravin} Y., {Kosowsky} A., 2009, \prd, 80, 023009

\bibitem[{{Kahniashvili} {et~al}\mbox{.}(2012){Kahniashvili}, {Maravin},
  {Natarajan}, {Battaglia}, \& {Tevzadze}}]{kahniashvili}
{Kahniashvili} T., {Maravin} Y., {Natarajan} A., {Battaglia} N., {Tevzadze}
  A.~G., 2012, ArXiv e-prints 1211.2769

\bibitem[{{Kahniashvili} \& {Ratra}(2005)}]{kahniashvili2}
{Kahniashvili} T., {Ratra} B., 2005, \prd, 71, 103006

\bibitem[{{Kandus}, {Kunze} \& {Tsagas}(2011){Kandus}, {Kunze}, \&
  {Tsagas}}]{kandus}
{Kandus} A., {Kunze} K.~E., {Tsagas} C.~G., 2011, \physrep, 505, 1

\bibitem[{{Kim}, {Olinto} \& {Rosner}(1996){Kim}, {Olinto}, \& {Rosner}}]{kim}
{Kim} E.-J., {Olinto} A.~V., {Rosner} R., 1996, \apj, 468, 28

\bibitem[{{Kim} {et~al}\mbox{.}(2007){Kim}, {Bolton}, {Viel}, {Haehnelt}, \&
  {Carswell}}]{kim2}
{Kim} T.-S., {Bolton} J.~S., {Viel} M., {Haehnelt} M.~G., {Carswell} R.~F.,
  2007, \mnras, 382, 1657

\bibitem[{{Kuhlen} \& {Faucher-Gigu{\`e}re}(2012)}]{2012MNRAS.423..862K}
{Kuhlen} M., {Faucher-Gigu{\`e}re} C.-A., 2012, \mnras, 423, 862

\bibitem[{{Meiksin}(2005)}]{Meiksin05}
{Meiksin} A., 2005, \mnras, 356, 596

\bibitem[{{Meiksin}(2009)}]{meiksinreview}
{Meiksin} A.~A., 2009, Reviews of Modern Physics, 81, 1405

\bibitem[{{Miralda-Escud{\'e}}(2003)}]{2003ApJ...597...66M}
{Miralda-Escud{\'e}} J., 2003, \apj, 597, 66

\bibitem[{{Paardekooper}, {Khochfar} \& {Dalla Vecchia}(2013){Paardekooper},
  {Khochfar}, \& {Dalla Vecchia}}]{2013MNRAS.429L..94P}
{Paardekooper} J.-P., {Khochfar} S., {Dalla Vecchia} C., 2013, \mnras, 429, L94

\bibitem[{{Pandey} \& {Sethi}(2013)}]{pandey}
{Pandey} K.~L., {Sethi} S.~K., 2013, \apj, 762, 15

\bibitem[{{Paoletti} \& {Finelli}(2012)}]{paoletti1}
{Paoletti} D., {Finelli} F., 2012, ArXiv e-prints 1208.2625

\bibitem[{{Paoletti}, {Finelli} \& {Paci}(2009){Paoletti}, {Finelli}, \&
  {Paci}}]{paoletti}
{Paoletti} D., {Finelli} F., {Paci} F., 2009, \mnras, 396, 523

\bibitem[{{Penprase} {et~al}\mbox{.}(2008){Penprase}, {Sargent}, {Martinez},
  {Prochaska}, \& {Beeler}}]{penprase}
{Penprase} B.~E., {Sargent} W.~L.~W., {Martinez} I.~T., {Prochaska} J.~X.,
  {Beeler} D.~J., 2008, in American Institute of Physics Conference Series,
  Vol. 990, First Stars III, {O'Shea} B.~W., {Heger} A., eds., pp. 499--503

\bibitem[{{Planck Collaboration} {et~al}\mbox{.}(2013){Planck Collaboration},
  {Ade}, {Aghanim}, {Armitage-Caplan}, {Arnaud}, {Ashdown}, {Atrio-Barandela},
  {Aumont}, {Baccigalupi}, {Banday}, \& et~al.}]{planck}
{Planck Collaboration} {et~al.}, 2013, ArXiv e-prints 1303.5076

\bibitem[{{Pogosian} {et~al}\mbox{.}(2011){Pogosian}, {Yadav}, {Ng}, \&
  {Vachaspati}}]{pogosian}
{Pogosian} L., {Yadav} A.~P.~S., {Ng} Y.-F., {Vachaspati} T., 2011, \prd, 84,
  043530

\bibitem[{{Schleicher} {et~al}\mbox{.}(2009){Schleicher}, {Galli}, {Glover},
  {Banerjee}, {Palla}, {Schneider}, \& {Klessen}}]{schleicher2}
{Schleicher} D.~R.~G., {Galli} D., {Glover} S.~C.~O., {Banerjee} R., {Palla}
  F., {Schneider} R., {Klessen} R.~S., 2009, \apj, 703, 1096

\bibitem[{{Shaw} \& {Lewis}(2012)}]{shaw}
{Shaw} J.~R., {Lewis} A., 2012, \prd, 86, 043510

\bibitem[{{Subramanian} \& {Barrow}(1998{\natexlab{a}})}]{1998PhRvD..58h3502S}
{Subramanian} K., {Barrow} J.~D., 1998{\natexlab{a}}, \prd, 58, 083502

\bibitem[{{Subramanian} \& {Barrow}(1998{\natexlab{b}})}]{subra}
{Subramanian} K., {Barrow} J.~D., 1998{\natexlab{b}}, Physical Review Letters,
  81, 3575

\bibitem[{{Sur} {et~al}\mbox{.}(2010){Sur}, {Schleicher}, {Banerjee},
  {Federrath}, \& {Klessen}}]{sur}
{Sur} S., {Schleicher} D.~R.~G., {Banerjee} R., {Federrath} C., {Klessen}
  R.~S., 2010, \apjl, 721, L134

\bibitem[{{Verner} \& {Ferland}(1996)}]{verner}
{Verner} D.~A., {Ferland} G.~J., 1996, \apjs, 103, 467

\bibitem[{{Wasserman}(1978)}]{wasserman}
{Wasserman} I., 1978, \apj, 224, 337

\bibitem[{{Widrow}(2002)}]{widrow}
{Widrow} L.~M., 2002, Reviews of Modern Physics, 74, 775

\bibitem[{{Yamazaki} {et~al}\mbox{.}(2012){Yamazaki}, {Kajino}, {Mathews}, \&
  {Ichiki}}]{yamazaki}
{Yamazaki} D.~G., {Kajino} T., {Mathews} G.~J., {Ichiki} K., 2012, \physrep,
  517, 141

\bibitem[{{Zahn} {et~al}\mbox{.}(2012){Zahn}, {Reichardt}, {Shaw}, {Lidz},
  {Aird}, {Benson}, {Bleem}, {Carlstrom}, {Chang}, {Cho}, {Crawford}, {Crites},
  {de Haan}, {Dobbs}, {Dor{\'e}}, {Dudley}, {George}, {Halverson}, {Holder},
  {Holzapfel}, {Hoover}, {Hou}, {Hrubes}, {Joy}, {Keisler}, {Knox}, {Lee},
  {Leitch}, {Lueker}, {Luong-Van}, {McMahon}, {Mehl}, {Meyer}, {Millea},
  {Mohr}, {Montroy}, {Natoli}, {Padin}, {Plagge}, {Pryke}, {Ruhl}, {Schaffer},
  {Shirokoff}, {Spieler}, {Staniszewski}, {Stark}, {Story}, {van Engelen},
  {Vanderlinde}, {Vieira}, \& {Williamson}}]{2012ApJ...756...65Z}
{Zahn} O. {et~al.}, 2012, \apj, 756, 65

\end{thebibliography}

\end{document}